# Effect of Substitution Group on Intramolecular Hydrogen Bond of Amino Alcohols from Raman spectroscopy

Honghui Zhao, Hongyuan Shen, Ao You, Yuanqin Yu*

School of Physics and Optoelectronic engineering, Anhui University, Hefei, 230601, China

* Author to whom correspondence should be addressed: yyq@ahu.edu.cn

**ABSTRACT**

Due to the simultaneous presence of two polar functional groups and flexible spatial structure, Aminoethanol (AE) is a model system for investigating the relationship between intramolecular hydrogen bonding and conformational equilibrium. In addition, Aminoethanol and their derivatives exhibit remarkable efficacy in the reversible capture of carbon dioxide. The intramoleculr hydrogen bond of 2-AE is determined by a subtle balance between electrostatic interactions, Van der Waals interactions, and steric effects. Changing the polarity of functional groups can regulate the strength of intramolecular hydrogen bonds. In this work, using spontaneous Raman spectroscopy combined with theoretical calculations, we investigated the effect of N-terminated substitution group on intramolecular hydrogen bond. When the H atom of NH2 functional group is replaced by electron-donating groups such as methyl and ethyl, it was observed experimentally that the red-shift of OH stretching vibration frequency caused by O-H... N intramolecular hydrogen bonding increases significantly and then the corresponding peak intensity increases. This indicates that with the introduction of substitutions on the N atom, the O-H... N intramolecular hydrogen bond in 2-AE is enhanced and the corresponding conformational population increases. The results of AIM and NCI analysis are consistent with experimental observations. These results provide insights for regulating the strength of intramolecular hydrogen bonds and also contribute to the strategy of CO2 capture.

**Keywords**：Amino ethanol（2-AE），Raman spectroscopy，Density functional method（DFT）

## 1. INTRODUCTION

Hydrogen bonding, a balance between strong covalent and weaker Van der Waals forces, involves hydrogen and electronegative elements like oxygen, nitrogen, or fluorine. It stems from the attraction between a polarized molecule's positive hydrogen and an electronegative atom's lone electrons in another molecule. Its strength relies on the electronegativity difference and the atoms' arrangement. Though less robust than covalent and ionic bonds, hydrogen bonds play a crucial role in determining molecules' 3D structure and geometry[1].

Vibrational spectroscopy, particularly through the red shift in XH stretching frequencies and enhanced transition intensity, serves as critical evidence for hydrogen bonding. This method quantitatively reflects bond strength and is essential for understanding molecular interactions, highlighting its significance in studying structure. The investigation employed FTIR and DFT to examine the impact of substituent groups on O-terminal carbon atoms in 2-AE on intramolecular hydrogen bond strength. It was found that electron-withdrawing groups increased, whereas electron-donating groups decreased hydrogen bond's strength. This article investigates the effects of electron-donating groups on the nitrogen atoms in AE on intramolecular hydrogen bonds[2]. Specifically, we explore the intramolecular hydrogen bond interactions in mono-methylaminoethanol（MMAE），dimethylaminoethanol（DMAE），and diethylaminoethanol（DEAE）. In the evaluation of intramolecular hydrogen bond strength, the widely used Atoms in Molecules (AIM) theory is employed, which can identify critical points in the distribution of electron density. Furthermore, this paper utilizes spontaneous Raman spectroscopy to obtain the N-H and O-H stretching spectra of the target molecules in carbon tetrachloride solution. Our simple expectation is that the substitution of H atom of NH2 group by the electron-donating -CH3 and -CH3CH2 group should increase the intramolecular hydrogen bond strength by making the NH2 group more alkali. Because the stronger the alkalinity of the functional group in a substance, the stronger its ability to act as a hydrogen bond acceptor to form hydrogen bonds with other hydrogen bond donors.

## 2. EXPRIMENT AND CALCULATE DETAIL

The foundational principles and experimental configuration of RAMAN SPECTROSCOPY have been thoroughly outlined in earlier research. In this context, a brief overview is provided. All experiments were carried out under standardized conditions, specifically at a controlled temperature of 25°C and a constant pressure of 1 atmosphere. To ensure the purity of the samples, alkanolamine samples were subjected to drying treatment using molecular sieves prior to the liquid-phase spontaneous spectroscopy experiments. In order to assure the reliability of the data, spectral collection was performed 15 times for each sample. Through meticulous data processing and error analysis, appropriate corrective actions were implemented, securing the accuracy and reproducibility of the experimental outcomes.

To validate experimental data and theoretically elucidate obtained RAMAN SPECTROSCOPY, this research utilized the Gaussian 16W [3]software package with the B3LYP/TZVP functional method, incorporating GD3BJ dispersion correction. The SMD implicit solvent model, chosen for its accuracy with halogen-bonded systems, simulated the solvent environment of ethanolamine and its derivatives in carbon tetrachloride[4]. Detailed conformational exploration was conducted in this solvent context. Initial geometric optimization and energy screening were performed using the molclus package's gentor tool and mopac's semi-empirical method, focusing on the five energetically lowest conformations. These were optimized with the B3LYP/TZVP method, and energy-ranked to identify the optimal conformation. This groundwork enabled precise Raman spectral and electron density calculations for the most energetically favorable conformation, thus offering a comprehensive analysis of the vibrational characteristics and distribution of dissolved molecules. This approach ensures the experimental results' accuracy and reproducibility, providing a robust theoretical framework for interpreting Raman spectroscopy data.

Following the aforementioned work, this project also employed the Reduced Density Gradient (RDG) method[5] within the Multiwfn software package for detailed topological analysis, taking into account the internal structure and performance parameters revealed by the computational model. The definition of RDG is:

$$s(r) = \frac{|\nabla \rho(r)|}{2(3\pi^2)^{\frac{1}{3}} \rho(r)^{4/3}} \quad (1)$$

Here, $s(r)$ represents the reduced density gradient at point $r$, $|\nabla \rho(r)|$ is the magnitude of the electron density gradient, and $\rho(r)$ is the electron density at point $r$. By correlating RDG values with the electron density and its second derivatives, a Non-Covalent Interaction (NCI) density plot can be generated. In the NCI plot, blue signifies attractive interactions, such as hydrogen bonds or Van der Waals forces, green usually represents weak attractions or nearly neutral interactions, possibly weak Van der Waals forces, and red indicates repulsive interactions, such as steric hindrance-induced repulsion. The shape and size indicate the spatial distribution and relative strength of the interaction zones. A broad distribution of the shape generally indicates stronger interactions, while smaller or more elongated shapes may suggest weaker interactions. In RDG diagram and RDG scatterplot, red represents steric effect, green represents van der Waals interaction, and blue represents strong electrostatic interaction.

Given the RDG isosurface method's limitations in finely analyzing intramolecular weak interactions, this project adopts an approach of precise quantification of the strength of intramolecular O-H…N hydrogen bonds at the BCP (Bond Critical Point) locations[6]. The formula for predicting the hydrogen bond system within this project is:

$$E_{\text{HB}} = -223.08 \cdot \rho(\text{BCP}) + 0.7423 \quad (2)$$

## 3. Result and Discussion

### 3.1 AIM and NCI analysis

Figure 1 illustrates the lowest-energy molecular conformations of four amino alcohols possessing OH...N intramolecular hydrogen bonds, along with AIM and RDG analysis diagrams computed within the SMD implicit solvent model in CCl4 based on these conformations.

The RDG method in AIM theory is used to quantify and visualize electrostatic interactions, van der Waals interactions, and steric effects in alcohol amine molecules. From the RDG scatter plot in Figure 1, it can be seen that the value of the spike characterizing intramolecular hydrogen bonding interactions is gradually decreasing according to the order of 2-AE, MMAE, DMAE and DEAE. This indicates that the strength of intramolecular hydrogen bonding is increasing as the substitution of H atom of NH2 group replaced by one -CH3, two -CH3 and two -CH3CH2 functional groups.

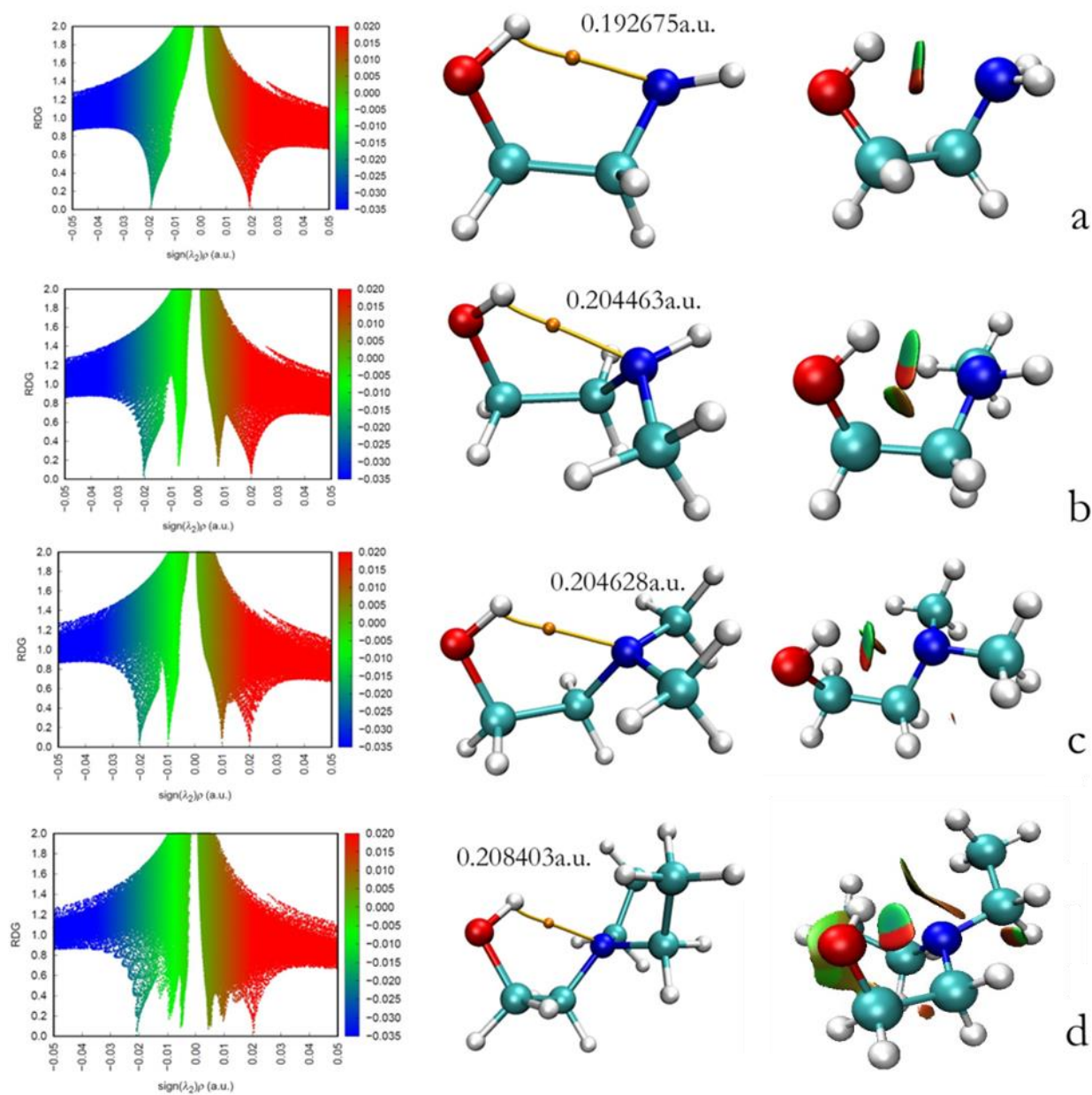

Figure1 RDG scatter plot, RDG plot, and AIM plot of the lowest energy conformations for four types of amino alcohols, 2-AE(a),MMAE(b),DMAE(c),DEAE(d).

Table 1 listed the distances between the hydrogen atom in the hydroxyl group and the nitrogen atom, the angles between the oxygen, hydrogen, and nitrogen atoms, the hydrogen bond energies calculated by BCP method from the electron density at critical points in Figure 1, and the Gibbs energy required to transition from the lowest energy conformation to the extended conformation in four types of amino alcohol molecules. Relative to 2-AE, the intramolecular hydrogen bond energy increases in MMAE, however, the increase in ΔG is not significant. Relative to MMAE, the intramolecular hydrogen bond energy in DMAE only shows a slight increase, while ΔG undergoes a significant change, increasing by 0.42 kcal/mol. DEAE exhibits only a slight increase in hydrogen bond energy and ΔG relative to DMAE. As the electron-donating group on the nitrogen atom increases, the angle formed by the oxygen, hydrogen, and nitrogen atoms approaches 180°. It is evident that when one -CH3 is substituted, the distance of OH⋯N decreases. However, from dimethyl substitution to diethyl substitution, the distance gradually increases. This is due to the increasing steric hindrance of the substituents, a conclusion supported by the RDG scatter plot. This indicates that the role of hydrogen bonding is much greater than the steric hindrance effect. In conclusion, the additional electron-donating group on the nitrogen atom has two main effects on the molecular conformation of amino alcohols: one is the strength of intramolecular hydrogen bonds, as evidenced by the abrupt change in hydrogen bond energy between MMAE and 2-AE molecules, and the other is the abrupt change in ΔG, as seen in the comparison of DMAE with 2-AE.

Table 1 The distance, angle, hydrogen bond energy, and ΔG of OH...N intramolecular hydrogen bonds in four types of amino alcohols.

| molecular | Distance (Å) OH…N | Angle (°) ∠OHN | Hydrogen bond energy (kcal/mol) | $\Delta G$ (kcal/mol) |
|---|---|---|---|---|
| 2-AE | 2.2299 | 115.83 | -3.5559 | -3.7456 |
| MMAE | 2.1960 | 117.35 | -3.8144 | -3.7651 |
| DMAE | 2.2031 | 117.53 | -3.8226 | -4.1861 |
| DEAE | 2.2249 | 117.62 | -3.9068 | -4.2959 |

**3.2 The Raman spectra of NH and OH stretching vibrations.**

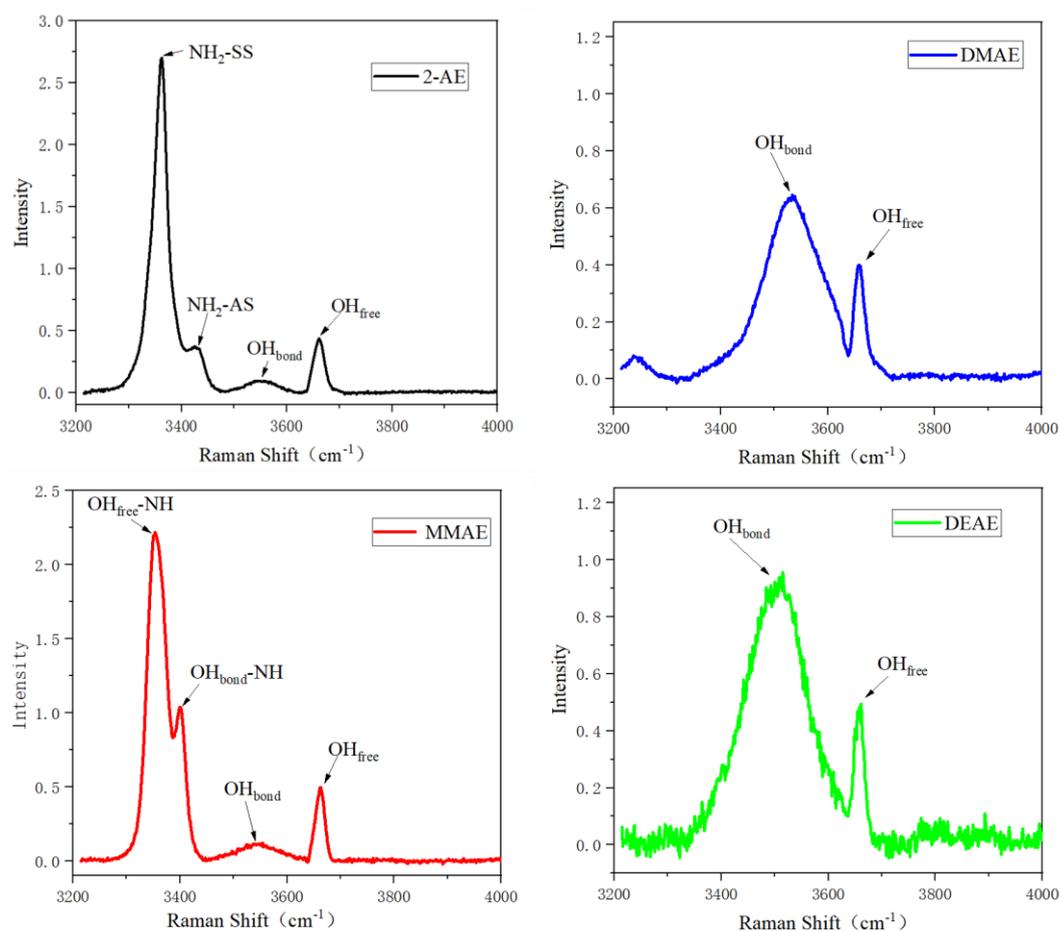

Figure 2 Spontaneous Raman spectra of 2-AE, MMAE, DMAE, DEAE in Carbon tetrachloride solution

From Figure 2, the vibration frequency of the free -OH conformer has not changed much. On the contrary, the vibration frequency of the hydrogen-bond conformer has been continuously reduced with the substitution of the substituent group. In combination with previous analysis, it can be seen that the strength of hydrogen-bond has been continuously enhanced with the substitution of the electron-donating group.

Table 2 listed a comparison between the calculated Raman spectra and the experimental Raman peaks of N-H and O-H c, along with their respective Raman frequencies. Subsequently, the observed Raman spectra are assigned based on the computational outcomes.

Table 2 Vibrational frequencies in the N-H and O-H stretching regions were observed and calculated for 2-AE, MMAE, DMAE, DEAE.

| molecular | $v_{obs}$ | $v_{cal}$ | Raman activity | Assignment |
|---|---|---|---|---|
| 2-AE | 3362 | 3352 | 160.8 | $NH_2$-SS |
|  | 3432 | 3433 | 80.7 | $NH_2$-AS |
|  | 3549 | 3591 | 61.2 | $OH_{bond}$ |
|  | 3661 | 3688 | 123.0 | $OH_{free}$ |
| MMAE | 3354 | 3356 | 107.6 | $OH_{free}$-NH |
|  | 3400 | 3400 | 127.9 | $OH_{bond}$-NH |
|  | 3541 | 3575 | 64.0 | $OH_{bond}$ |
|  | 3662 | 3688 | 138.0 | $OH_{free}$ |
| DMAE | 3528 | 3576 | 69.3 | $OH_{bond}$ |
|  | 3659 | 3710 | 117.6 | $OH_{free}$ |
| DEAE | 3500 | 3573 | 66.4 | $OH_{bond}$ |
|  | 3658 | 3710 | 126.7 | $OH_{free}$ |

Due to inherent approximations and errors in theoretical calculations, a correction factor of 0.9461001 was employed for the computed Raman spectra in this investigation.

Compared to 2-AE, MMAE exhibits only a slight redshift and a slight increase in peak height for the Raman peak representing intramolecular hydrogen bonds ($OH_{bond}$), indicating a slight increase in the strength of intramolecular hydrogen bonds and the proportion of conformations with intramolecular hydrogen bonds in carbon tetrachloride solution. However, relative to MMAE, DMAE shows a significant increase in the peak intensity of the Raman peak representing intramolecular hydrogen bonds, along with a larger redshift. This indicates a greater increase in the strength of intramolecular hydrogen bonds and a larger proportion of conformations with intramolecular hydrogen bonds in DMAE compared to MMAE. The reason for this is that the electron-donating effect of the dimethyl substitution is more pronounced than that of the methyl substitution, leading to a greater enhancement of intramolecular hydrogen bond strength. As a result, the hydrogen bonds are more difficult to break, leading to a greater increase in the proportion of conformations with intramolecular hydrogen bonds. This finding is consistent with the DFT computational results. As mentioned earlier, as the hydrogen atom on the nitrogen atom is substituted by an electron-donating group, the strength of intramolecular hydrogen bonds becomes stronger, and the proportion of conformations with intramolecular hydrogen bonds increases in carbon tetrachloride solution. Additionally, electron-donating effects induced by symmetric substitutions are much stronger than those induced by asymmetric substitutions.

## 4. Conclusion

We investigated the impact of substituting electron-donating groups for hydrogen atoms on the functional groups of amines on intramolecular hydrogen bonding. Specifically, we measured the RAMAN SPECTROSCOPY of 2-AE, MMAE, DMAE, and DEAE, and computed and analyzed critical point electron densities to assess the strength of intramolecular hydrogen bonds. The analysis of RAMAN SPECTROSCOPY, NCI, and BCP results indicated that, with the substitution of electron-donating groups, the frequency of intramolecular hydrogen bonds decreased while their strength increased. Additionally, the electron-donating effect induced by symmetric substitutions was significantly stronger than that induced by asymmetric substitutions. The primary reason for the increase in hydrogen bond strength caused by electron-donating groups is their ability to enhance the alkalinity of the N atom, thereby better attracting H atoms. Consequently, the energy required to break the hydrogen bonds increases, rendering the hydrogen bonds more stable.

Finally, this study demonstrates that the substitution of electron-donating groups for H atoms at the N-terminal of amines can enhance the strength of intramolecular hydrogen bonds. This provides a new theoretical basis for the modification and optimization of amino alcohols, which may assist other researchers in designing more suitable tools for investigating hydrogen bonding interactions.